# Vers une architecture d'intégration sémantique des composants métier

# Towards an architecture for semantic integration of business components

*Hicham ELASRI* [1, 3]
**E-mail :** h.elasri@active3d.net

*Larbi KZAZ* [1, 2]
**E-mail :** larbi_kzaz@yahoo.fr

*Abderrahim SEKKAKI* [1]
**E-mail :** a.sekkaki@fsac.ac.ma

(1) Université Hassan II - Ain chock
Faculté des Sciences Ain Chock Casablanca Maroc.
Laboratoire d'Informatique et d'Aide á la Décision (LIAD).
Equipe Architecture et Gestion des Systèmes Distribués (AGSD).
B.P 5366 Maarif Casablanca 20100 Maroc
Tél : +212 522 23 06 80 / 84
Fax : +212 522 23 06 74.

(2) Institut Supérieur de Commerce et d'Administration des Entreprises (ISCAE)
Km 9,500 Route de Nouasseur BP. 8114 - Casablanca Oasis
Tél. : + 212 5 22 33 54 82 à 85
Fax : +212 522 33 54 96

(3) GROUPE ARCHIMEN / ACTIVE3D/Casablanca
40 bld D'Anfa 200 Casablanca
Tél. +212 5 22 22 62 79
Mob. 0033 6 83 69 09 52

**Résumé**

Aujourd'hui, les composants réutilisables sont disponibles dans plusieurs référentiels. Ces derniers sont certes conçus pour la réutilisation. Cependant, cette réutilisation n'est pas immédiate ; elle nécessite, en effet, de passer par quelques opérations conceptuelles essentielles, dont notamment, la recherche, l'intégration, l'adaptation et la composition. Nous nous intéressons dans le présent travail au problème de l'intégration sémantique des Composants Métier (CM) hétérogènes. Ce problème a été souvent posé en termes syntaxiques alors que le véritable enjeu est en fait d'ordre sémantique. Notre contribution porte sur la proposition d'une architecture d'intégration des CM, ainsi qu'une méthode de résolution des conflits sémantiques de nommage rencontrés lors de l'intégration des CM.

**Mots clefs :**
Composant métier, Intégration sémantique, Ontologie, Web sémantique.**.**

**Abstract**

Today, reusable components are available in several repositorys. These are certainly conceived for re-use. However, this re-use is not immediate, it requires, in effect, to pass by some essential conceptual operations, among which in particular, research, integration, adaptation, and composition. We are interested in the present work to the problem of semantic integration of heterogeneous Business Components. This problem is often put in syntactical terms, while the real stake is of semantic order. Our contribution concerns an architecture proposal for Business components integration and a resolution method of semantic naming conflicts, met during the integration of Business Components.

**Key-words:**
Business Component, Semantic Integration , Ontology, Semantic Web.

# Introduction

L'approche à base de composants est considérée depuis les années 90, comme un paradigme de développement des systèmes d'information (Barbier, 2002). Elle a pour but de réduire de façon conséquente les coûts et les délais de mise à disposition des logiciels. Le principe de cette approche consiste à construire de nouveaux systèmes à partir de composants disponibles. Le recours à cette approche dès les premières phases de développement présente un réel intérêt (oussalah 2005). Deux types de problématique de recherche existent dans le domaine de la réutilisation. Celles qui relèvent de l'ingénierie pour la réutilisation (« design for reuse ») visent à développer des méthodes et des outils pour supporter le processus de production de composants. Celles qui relèvent de l'ingénierie par réutilisation (« design by reuse ») ont pour objectif de proposer des méthodes et des outils pour exploiter des composants réutilisables (Puljate, 04). Notre recherche s'inscrit dans le cadre de la conception des SI à base de composants métier réutilisables. Elle concerne plus précisément la problématique de l'intégration sémantique des composants métier (CM). Plusieurs types de conflit sémantique doivent être résolus lors de l'intégration des CM ; nous nous intéressons aux conflits de nommage dans les CM. Nous proposons dans ce cadre une architecture ainsi qu'une méthode d'intégration basées sur les ontologies, ces dernières étant utilisées pour des questions similaires du Web sémantique, et sur le calcul de la mesure de similarité pour la résolution des conflits de nommage dans les CM. Dans la section 1 nous présentons le paradigme de Composant Métier; dans la section suivante nous abordons la problématique de l'intégration sémantique des CM, les différents types de conflit qu'elle soulève, puis nous nous intéressons aux ontologies et à leur utilisation dans le Web sémantique, et proposons d'étendre leur application aux CM. Dans la section 3 nous présentons une architecture ainsi qu'une méthode d'intégration sémantique des CM fondée sur les ontologies. Nous poursuivons cette section par des exemples d'application illustrant la méthode. En fin d'article nous exposons les différentes perspectives d'extension de notre solution.

# 1. Le paradigme Composant Métier.

Le terme composant est largement utilisé dans le domaine de la réutilisation, avec une connotation générale d'entité autonome réutilisable. Le développement des SI à base de composants consiste à construire le SI à partir d'un ensemble de composants disponibles. Le paradigme composant métier, quant à lui, se base sur une catégorie particulière de composants appelée composant métier (CM).

## 1.1 Définition

Dans la littérature, on trouve plusieurs définitions du concept de composant métier. Nous retenons les deux définitions suivantes : la première est celle donnée par Herzum et Sims dans (Herzum, 2000), "*A business component is the software implementation of an autonomous business concept or business process. It consists of all the software artefacts necessary to represent, implement, and deploy a given business concept as an autonomous, reusable element of a larger distributed information system*". La deuxième est celle donnée par F. Barbier (Barbier, 2002) « *A business component models and implements business logic, rules and constraints that are typical, recurrent and comprehensive notions characterizing a domain or business area* ».

Un CM modélise donc et implémente une entité significative par rapport à un métier de l'entreprise. Il capture et décrit dans des termes issus du vocabulaire de l'entreprise et de son métier, des concepts, des événements ou des processus.

Selon le paradigme CM, un SI d'entreprise est construit à partir d'un ensemble de CM pouvant émaner de différentes sources. Le SI commercial d'une entreprise, par exemple, pourrait être conçu à partir de CM tels que : {«Vente », « Produit », « Client » etc..}.

## 1.2 Classification

La classification des composants métier (CM) peut être fondée sur le « type de connaissances » qu'ils représentent. Selon (Herzum, 1999) un CM peut représenter soit une entité : CM-entité (des personnes, des clients, des fournisseurs, des adresses postales, des factures, etc.), soit un processus : CM-processus (processus de d'achat, processus de vent, etc.) soit un utilitaire : CM-utilitaire (Note, Code, etc). A ces trois catégories. (Hassine, 2005) a ajouté une quatrième, il s'agit de la catégorie donnée : CM-donnée. Notons que les deux dernières catégories : CM-utilitaire et CM-donnée, sont de granularité faible, et ne sont pas destinés à être réutilisées indépendamment des composants des deux autres catégories ; ils leurs servent de base dans la conception. Par contre, les CM-processus et CM-entité, qui sont de granularité forte, peuvent être réutilisés indépendamment. De ce qui précède, nous pouvons déduire l'existence d'une certaine hiérarchisation entre les CM. Les CM-processus se situent au niveau le plus élevé ; ils se basent sur les CM-entité situés au niveau suivant. Les CM-Utilitaire se trouvent au niveau immédiatement inférieur, suivis des CM-donnée qui sont placés au niveau le plus bas de la hiérarchie (Barbier, 02).

Une autre classification consiste à distinguer entre les « composants verticaux », réutilisables uniquement au sein d'un même domaine, et les « composants horizontaux », réutilisables dans différents domaines (Oussalah 05).

Par ailleurs, notons qu'il y'a lieu de distinguer entre les aspects conceptuel et logiciel du concept de CM (Khayati, 2005) et (Barbier 2002). Les composants logiciels sont décrits par leur codage dans un langage de programmation donné et pour une infrastructure donnée. Les composants conceptuels sont quant à eux décrits dans des langages de modélisation standards et technologiquement neutres tel que UML. Le développement à base de composants peut alors se définir comme la réutilisation et l'intégration des modèles décrivant les composants (Vauttier, 2003). Dans la suite de ce travail, le terme de CM désignera l'aspect conceptuel. Cet aspect étant fondamental dans les activités de spécification et de conception des SI à base de CM.

## 2. L'intégration sémantique.

### 2.1 Définition

Le terme interopérabilité sémantique a été introduit en 1995 par Sandra Heiler (Heiler, 1995); elle lui a donné la définition suivante : *« Interoperability among components of large-scale distributed systems is the ability to exchange services and data with another. (…) Semantic interoperability ensures that these exchanges make sense – that the requester and the provider have a common understanding of "the meanings" of the requested services and data. »*,(Barbier, 2002), (Mellal, 2007), (Vallecillo, 2000), (Vernadat, 2007).

L'interopérabilité sémantique est la capacité des composants à échanger des données et des services tout en partageant le sens des échanges. L'interopérabilité sémantique permettant ainsi l'intégration sémantique. L'intégration sémantique des composants passe donc nécessairement par la détection et la résolution des conflits sémantiques pouvant exister entre composants.

### 2.2 Les conflits d'intégration

L'échange de données et de services entre CM peut donner lieu à différents types de conflit sémantique. Plusieurs chercheurs, (Hendriks, 2007) ,(Kavouras , 2004) ,(Izza, 2006),  (Visser, 2001),ont identifié trois types de conflits sémantiques: conflit de confusion, conflit de mesure et conflit de nommage.

#### 2.2.1 Conflit de confusion

(Hendriks, 2007)a défini le conflit de type confusion comme suit : *« Confounding conflicts: information items appear to have the same meaning but differ in reality due to e.g. a different temporal context (e.g. 'occurred 5 minutes ago') »*

Le conflit de type confusion est donc lié aux données contextuelles ayant les mêmes apparences, mais changeant de comportement par rapport au temps. Par exemple cette personne est un *employé*, hier a été *responsable*, aujourd'hui est devenue *directeur*. Donc elle est toujours *employé*, mais son grade a changé. Cet exemple est schématisé dans la figure ci-dessous par le composant métier « employé »

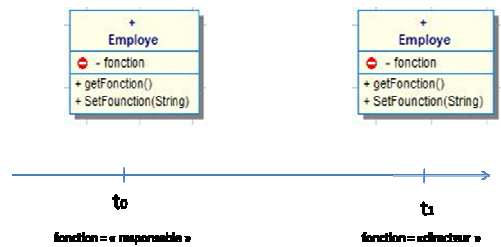

**Figure 1 : Exemple illustratif de conflit de confusion**

#### 2.2.2 Conflit de mesure

Ce type de conflit survient quand deux systèmes expriment la même valeur à l'aide d'unités différentes (Hendriks, 2007), (Izza, 2006), (visser, 2001). Par exemple si on veut intégrer deux composants métier « Produit », dont l'attribut « Prix » du premier a pour unité de mesure l'Euro et l'attribut du deuxième composant utilise le dollar (cf. figure 2).

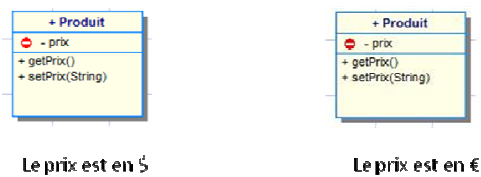

**Figure 2 : Exemple illustratif du conflit de mesure**

#### 2.2.3 Conflit de nommage

(Visser, 2001) définit le conflit de type nommage comme suit : *«Naming conflicts occurs when naming schemes of information differ significantly.A frequent phenomenon is the presence of homonyms and synonyms »*.

Les conflits de nommage sont dus à la présence d'homonymes ou de synonymes.

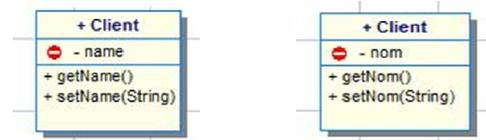

**Figure 3 : exemple illustratif du phénomène synonymie**

Nous nous intéresserons par la suite exclusivement au conflit de nommage.

### 2.3 Les mécanismes d'intégration

Les mécanismes d'intégration ont pour objectif de résoudre les conflits dus à l'hétérogénéité des CM, dans le but de les rendre interopérables. Dans la littérature on peut distinguer deux catégories de mécanismes d'intégration : les mécanismes qui se basent sur des modèles préétablies (les modèles de composant) et les mécanismes qui se basent sur les ontologies. Notre proposition s'appuie sur les ontologies, comme élément clé pour assurer la résolution des conflits.

### 2.3.1 Les ontologies.

Les ontologies proposent une compréhension commune et partagée d'un domaine, tant au niveau des utilisateurs humains qu'au niveau des applications logicielles. Elles sont devenues un outil clé dans la représentation des connaissances; leurs applications sont nombreuses et font l'objet d'intenses travaux dans différents domaines : intelligence artificielle, traitement de la langue naturelle, la recherche d'information, le travail collaboratif etc. (Amar, 2007).

Selon (Pujalte, 2004), une ontologie de domaine est définie par deux éléments complémentaires :

**Le modèle de domaine :** Le modèle de domaine est composé de concepts et de liens entre ces concepts. Il contient les concepts contenus dans les ontologies locales (ontologies sources) et introduit une taxonomie (hiérarchisation) de ces concepts.

**Le thésaurus :** contient les termes dérivés et les définitions des concepts définis dans le modèle de domaine. Il fournit aussi le vocabulaire générique permettant de décrire des contextes. Les termes dérivés sont les synonymes et homonymes des concepts.

L'alignement des ontologies (recherche de mappings, appariement ou mises en correspondance) est une tâche particulièrement importante dans les systèmes d'intégration, ce thème a donné lieu à de nombreux travaux (Safar, 2009). Nous nous appuyons sur les résultats de ces travaux pour réaliser l'intégration sémantique des CM. Les CM représentent dans notre cas les ressources à intégrer par le biais des ontologies.

Notre recours aux ontologies de domaine peut être justifié pour diverses raisons : Tout d'abord, une ontologie de domaine porte, par définition, sur les concepts relatifs à un domaine applicatif particulier ; ceci correspond parfaitement à notre problématique, puisque la conception d'un SI vise généralement un domaine métier. Ensuite, Les ontologies de domaine sont réutilisables à l'intérieur d'un même domaine (Benayahe, 2005) (Lando, 2006) cette propriété est très intéressante pour envisager la réutilisation verticale des CM, qui est l'objectif central des approches par composant.

### 2.3.2 Le Web sémantique.

Selon le W3C, *« le Web sémantique est une vision : l'idée est que les données sur le web soient définies et liées de manière à pouvoir être utilisées par des machines sur le web non seulement pour des fins d'affichage, mais pour l'automatisation, l'intégration et la réutilisation sur des plates-formes variées »*.

Le Web sémantique est une extension du Web actuel donnant un sens au contenu. Il s'appuie entre autres sur des ontologies et des langages, pour attribuer et représenter le sens de ses ressources, et ce afin de permettre à des programmes et à des agents d'y accéder à l'aide de langages développés par le W3C.

Dans notre cas le Web sémantique peut servir de plateforme aux concepteurs pour la recherche, l'identification de composants réutilisables nécessaires à la conception d'un système d'information. En effet, plusieurs travaux portent sur le processus de recherche de composants réutilisables. Ceux-ci visent à faciliter la réutilisation des composants au moyen d'approches plus ou moins naturelles dont le but est de fournir des composants adaptés aux besoins des concepteurs de systèmes d'information (Pujalte, 2004).

## 3. Une proposition pour l'intégration sémantique des CM.

### 3.1 Définition de l'intégration des CM

Les CM fournissent des services et ou des données, l'intégration sémantique des CM consiste à attribuer du sens aux données et aux services fournis afin d'assurer les échanges des données et des services entre CM hétérogènes.

### 3.2 Architecture d'intégration.

L'architecture d'intégration que nous proposons exploite les résultats de certains travaux menés sur les composants et les ontologies :

- La transformation en ontologies des CM décrits dans un langage de modélisation tel que UML.
  Cette transformation est rendue possible grâce notamment aux travaux de (Gasevic 2004) qui a présenté une approche basée sur le langage XSLT, pour la génération automatique de OWL à partir d'UML.
- L'alignement des ontologies résultat des transformations des CM en ontologies, à l'aide d'une ontologie de domaine.
  Cette méthode est assimilable aux méthodes d'alignement des ontologies s'appuyant sur des ressources complémentaires ciblées, appelés ontologies de support ou de background (Sabou, 2006), (Safar, 2009) et (Aleksovski, 2006a). Dans notre cas l'ontologie relatif au domaine du SI à concevoir et dont sont issus les CM à intégrer, joue le rôle de ressource complémentaire ciblée et sera donc l'ontologie de support.

Pour illustrer l'architecture que nous proposons, nous supposons disposer de deux systèmes d'information S1 et S2 à base de composants, qu'on voudrait intégrer sémantiquement, afin d'avoir un nouveau SI élaboré à partir des composants des deux systèmes S1 et S2. S1 possède un ensemble ($ECM^{s1}$) des $CM^{s1}_1....CM^{s1}_n$ et S2 possède un ensemble ($ECM^{s2}$) des $CM^{s2}_1........CM^{s2}_p$.

Nous avons donc un ensemble $ECM^{S1S2}$ qui représentera l'union des deux ensembles. Par la suite, tous les éléments de l'ensemble $ECM^{S1S2}$ sont candidats à l'intégration. Nous supposons que les taches d'identification, de recherche, de sélection des CM et la phase de la transformation des CM, supposés être représentés en UML, vers des ontologies ($O_{n+p}$) sont réalisés. Cette transformation étant possible (Gasevic, 2004), nous disposons alors d'un ensemble d'ontologies $(EOCM)(OCM_1……OCM_{n+p})$ produites à partir de $ECM^{S1S2}$.

L'intégration sémantique des composants métier se déroulera selon les étapes suivantes :

1. Soit $ECM^{S1S2}$ l'ensemble des composants métier candidats à l'intégration.
2. L'ensemble $ECM^{S1S2}$ est transformé en un autre ensemble d'ontologies ($EOCM^{S1S2}$)
3. La TIO reçoit l'$EOCM^{S1S2}$ et par suite décide de la technique d'intégration d'ontologies à adopter, il s'agit de la technique d'alignement dans notre cas.
4. Détection et résolution des conflits sémantiques par la méthode de mesure de la similarité.
5. Création de l'ontologie de représentation par la correspondance entre les concepts des ontologies produites à partir des composants métier et Obtention d'un CM résultat : $CM_R$ qui représente les connaissances des composants métier qui ont subi le traitement d'intégration.

Le schéma ci-après présente l'architecture générique d'intégration des CM que nous proposons.

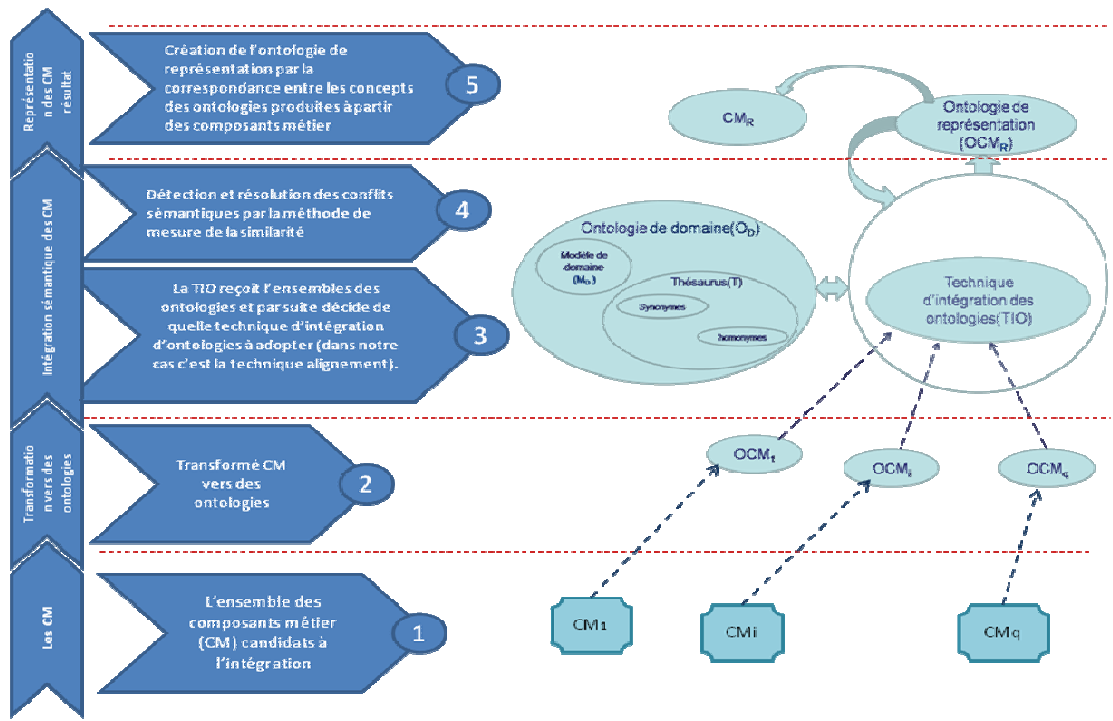

**Figure 4: Le processus d'intégration sémantique des CM.**

L'architecture d'intégration des CM comporte plusieurs éléments :

**L'ontologie de domaine :** Elle modélise les connaissances relatives au domaine du système d'information, objet de l'intégration. Nous exploiterons la réutilisation verticale des ontologies de domaine. Selon (Puljate, 2004) : Une ontologie de domaine $O_D$ est composée de deux éléments complémentaires, le modèle de domaine $M_D$, et le thésaurus (T) ; elle sera notée par la suite $O_D = (T, M_D)$

**Technique d'intégration des ontologies (TIO) :** c'est dans cet élément de notre architecture que se réalisent les traitements sur les ontologies ($OCM_1…OCM_q$) produites à partir des CM. Il existe plusieurs techniques d'intégration des ontologies : les techniques de transformation qui ne traitent qu'une seule ontologie, les techniques de 'mapping' et 'fusion', conçues pour ne traiter que deux ontologies. Nous n'avons retenu aucune de ces deux techniques dans la mesure où nous sommes dans un environnement multi-ontologies, la seule technique d'intégration qui nous semble adaptée à notre cas est la technique d'alignement des ontologies. Cette dernière permet de faire plusieurs types de correspondance entre les concepts (un à un, un à plusieurs, plusieurs à un et plusieurs à plusieurs).

**Ontologie de représentation :** l'ontologie résultat de l'alignement est notée ($OCM_R$). Elle permet d'avoir :

- Un $CM_R$ représenté par exemple en UML, afin d'aider les concepteurs et les architectes des SI, à élaborer leurs systèmes.
- Un $OCM_R$ qui peut devenir une source ontologique d'un autre traitement ou d'une autre itération d'intégration.

### 3.3 Mesure de la similarité syntaxique.

Nous proposons dans cette section une méthode de mesure de la similarité syntaxique entre concepts ; cette méthode sera utilisée par la méthode de mesure de la similarité sémantique développée dans la prochaine section.

*Soit σ' la méthode de calcul de la similarité*

Soit Sc : l'ensemble des concepts

$\forall$ *Ci* $\in$ *Sc, Ci est défini par le couple (Tri, EDi), avec Tri : le terme désignant le concept Ci et EDi l'ensemble de ses définitions.*

**σ' : Sc × Sc → {0, 1}**

*Soient C1, C2 deux concepts de Sc*

*Deux cas sont à distinguer :*

**Cas 1** : C1 et C2 sont des concepts atomiques.

*Si   Tr1 =Tr2    alors    σ' (C1, C2) = 1*

*Sinon   σ' (C1, C2) = 0*

***Fsi***

**Cas 2** : C1 et C2 sont composites.

*C1 et C2 s'écrivent alors C1 = (C11.., C1i,…., C1n) et C2 = (C21 …., C2j ,…., C2n)*

*La méthode σ' est déterminée ainsi :*

*σ' (C1, C2) =1/n ($\sum_{i,j}$ σ'(C1i, C2j))   1 <= i,j <= n*

La méthode σ' proposée, prend la valeur 1 lorsque les concepts sont syntaxiquement identique et 0 dans le cas contraire.

### 3.4 Mesure de la similarité sémantique.

La méthode de mesure de la similarité sémantique entre concepts, est basée sur les ontologies de domaine et sur la méthode de mesure de la similarité syntaxique σ', définie ci avant.

Notons σ la méthode de détermination de la similarité sémantique entre concepts, σ est définie comme suit :

*σ : Sc × Sc → {0, 1},*

Soient C1 et C2 deux concepts de Sc, $O_D$ l'ontologie de domaine, et T le thésaurus

**Cas 1** : C1 et C2 appartiennent à $O_D$ et C1 et C2 admettent un terme dérivé dans T:

Si le terme dérivé est synonyme

alors   σ (C1, C2) = 1

Sinon

Si le terme dérivé est homonyme

alors   σ (C1, C2) = 0

Fsi.

Fsi.

**Cas 2** : (C1 et C2 appartiennent à $O_D$ et C1 et C2 n'admettent pas de terme dérivé dans T) ou (C1 et C2 n'appartiennent pas à $O_D$):

σ (C1, C2) = σ' (C1, C2)

La méthode *σ* ainsi proposée, retourne la valeur 1 lorsque les concepts sont synonymes et 0 s'ils sont homonymes ; permettant ainsi de détecter et de résoudre les conflits sémantiques entre CM.

Le tableau ci-après résume les différents cas de figure pouvant se produire entre deux concepts C1 et C2.

|  | Egalité sémantique entre C1 et C2 | | Egalité syntaxique entre C1 et C2. | |
|---|---|---|---|---|
|  | C1 et C2 appartiennent à $O_D$ | | | |
|  | Il existe dans T un terme dérivé synonyme entre C1 et C2 | Il existe dans le T un terme dérivé homonyme entre C1 et C2. | C1 et C2 n'appartiennent pas à $O_D$ | |
| Valeurs de mesure de similarité | *σ(C1, C2)=1* | *σ*(C1 , C2)=0 | *σ'*(C1 , C2)=1 | *σ'*(C1 , C2)=0 |
|  | C1 et C2 sont synonymes | C1 et C2 sont homonymes | C1 et C2 sont égaux | C1 et C2 sont différents |

**Tableau 1 : les valeurs possibles des mesures syntaxique et sémantique**

## 3.5 Exemples d'application.

Nous proposons ci-après trois illustrations permettant d'une part de valider notre solution et d'autre part de l'expliquer.

**Illustration 1 :**

Soit $CM_1$ et $CM_2$ deux composants synonymes appartenant respectivement à deux systèmes S1 et S2, notre méthode se déroulera comme suit :

- Détermination des ontologies ($OCM_1$ et $OCM_2$) à partir des deux composants.

**Cas 1 : Composants synonymes.**

- Calcul de la similarité : étant donnée que $CM_1$ et $CM_2$ sont synonymes, on aura pour chaque couple d'éléments (ep $CM_1$) de $CM_1$ et (eq $CM_2$) de $CM_2$ avec leurs concepts ontologiques respectifs (ep $OCM_1$) (eq $OCM_2$), σ (ep $CM_1$, eq $CM_2$) = 1. La similarité globale sera donc calculée ainsi : σ ($OCM_1$, $OCM_2$) = $\sum_{i\,j}$ σ (ei $CM_1$, ej $CM_2$ ))/n  1≤i≤n , 1≤j≤n  Donc σ ($CM_1$, $CM_2$) = (n /n)  =1 ;

Il en découle d'après notre méthode que $CM_1$ et $CM_2$ sont synonymes, ce qui confirme notre hypothèse.

**Cas 2 : Composants homonymes.**

- Calcul de la similarité : étant donnée que $CM_1$ et $CM_2$ sont homonymes, il doit exister au moins deux éléments (ep $CM_1$) de $CM_1$ et (eq $CM_2$) de $CM_2$ avec leurs concepts ontologiques respectifs (ep $OCM_1$) (eq $OCM_2$), tel que ep $OCM_1$ est différent de eq $OCM_2$ et par conséquent σ (ep $CM_1$, eq $CM_2$) = 0. La similarité globale sera donc calculée ainsi : σ ($OCM_1$, $OCM_2$)= (σ (ep $CM_1$ , eq $CM_2$ ) + $\sum_{ij}$ σ (ei $CM_1$ , ej $CM_2$ ))/n   avec ( 1≤i,j≤n et (i,j) ≠ (p,q) ) Donc σ ($OCM_1$, $OCM_2$) = (0  + $\sum_{ij}$ σ (ei $CM_1$ , ej $CM_2$ )/n)  ≠1 ;

Il en découle que $CM_1$ et $CM_2$ ne peuvent être synonymes ; ils sont homonymes, ce qui confirme notre hypothèse.

**Illustration 2 :**

Supposons que nous avons deux composants métier : le premier noté $CM_1$ : client (nom, âge) et le deuxième $CM_2$ : client (nom, prénom).

Les deux composants ont le même terme qui les désignent client = client ; ils ont la même apparence et par conséquent σ' (CM1, CM2) =1;

Si les concepts associés aux deux composants appartiennent à l'ontologie de domaine et si le thésaurus contient des termes dérivés entre ces deux concepts, donc ces deux concepts sont homonymes et la méthode de similarité σ aura la valeur 0. Dans ce cas, nous pouvons conclure que nous avons un conflit de type nommage puisque nous avons les deux concepts ayant la même apparence et une valeur de similarité (σ) égale à 0.

Dans le cas où ces concepts n'appartiennent pas à l'ontologie de domaine, nous vérifierons leurs sous concepts : nom, prénom et âge.

Puisque ces sous concepts atomiques alors σ (nom, nom) = σ' (nom, nom) = 1, et  σ (prénom, age) = σ' (prénom, age) =0.

Par conséquent σ (CM1, CM2) = ½ (σ(nom, nom)  + σ ( prénom, âge)) et { σ(nom, nom)  = σ'(nom, nom)  , σ ( prénom, âge)= σ' ( prénom, âge)} = ½ (1 + 0 )  ≠1 donc σ(CM1, CM2 ) =0. Ces deux concepts ont donc la même apparence mais avec une valeur de similarité sémantique égale à 0, par conséquent ces concepts présentent un conflit de type nommage.

**Illustration 3 :**

Considérons le cas de la fusion de deux bibliothèques. A travers ce cas, nous illustrons notre méthode de résolution des conflits sémantiques pouvant apparaître lors de la fusion des deux bibliothèques. Les figures 5 et 6 sont des représentations UML des diagrammes des classes de chacune des deux bibliothèques. Nous supposons disposer des résultats de la transformation en ontologies des composants des deux bibliothèques

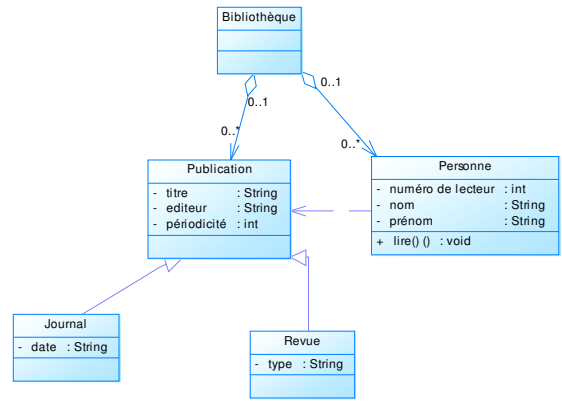

**Figure 5 : Diagramme de classes de bibliothèque 1**

La bibliothèque représentée dans la figure 5 comporte les journaux et les revues scientifiques. Elle traite la consultation en ligne des publications, c'est-à-dire que la personne peut lire en ligne une publication qui est décrite par son titre, son éditeur et sa périodicité, cette publication peut être soit un journal décrit  par sa date de sortie, soit une revue.

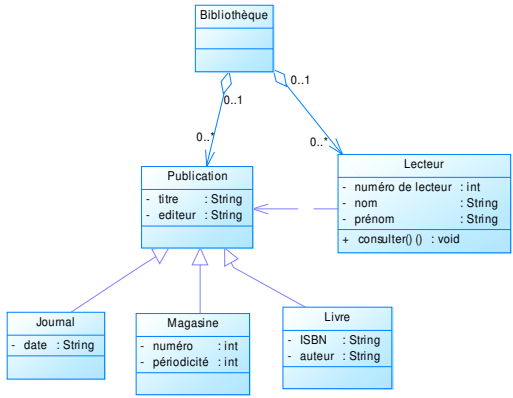

**Figure 6: Diagramme de classes de bibliothèque 2**

La bibliothèque représentée dans la figure 6 comporte les journaux, les livres et les magazines. Elle traite la consultation en ligne des publications, c'est-à-dire que le lecteur peut lire en ligne une publication décrite par son titre et son éditeur cette publication qui peut être soit un journal qui est représenté par sa date de sortie, soit un livre décrit par son code ISBN et son auteur, soit un magasine décrit par son numéro et sa périodicité.

A chacun des diagrammes de classe est associé un diagramme des composants (Figure 7,8).

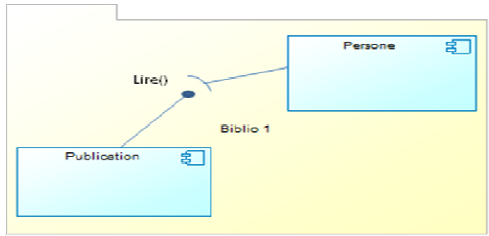

**Figure 7 : Digramme de composants de bibliothèque 1**

La figure 7 représente le diagramme de composants de bibliothèque 1, cette dernière comporte deux composants métier de type entité : « Personne » et « Publication ». L'interface du premier fournit l'opération « lire () » requise par le second.

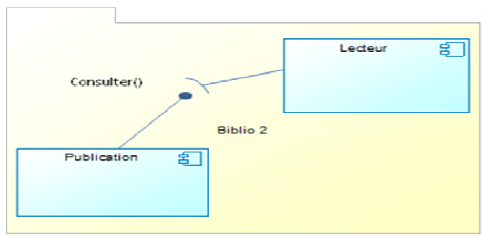

**Figure 8 : Digramme de composants de bibliothèque 2**

Le diagramme de composant de bibliothèque 2, figure 8, comporte lui aussi deux CM de type entité : « Lecteur » et « Publication ». L'interface du premier fournit l'opération « Consulter () » requise par le second.

L'intégration sémantique des quatre composants métier se déroulera comme suit :

1. Identification des composants métier « Personne » et « Publication » de « Biblio1 » et les composants métier « Lecteur » et « Publication » de « Biblio2 ».

2. Production des ontologies relatives aux composants métier : OntoPersonne, OntoPublication, OntoLecteur et OntoPublication.

3. L'environnement qui implémente TIO reçoit en entrée les quatre ontologies.

4. Calcul de la mesure de similarité appliquée sur les ontologies et leurs sous-concepts σ (OntoPersonne, OntoLecteur) et σ (OntoPublication, OntoLecteur).

Le tableau ci-après détaille le calcul des mesures de similarité entre les sous concepts des deux ontologies :OntoPersonne et OntoLecteur

| OntoPersonne \ OntoLecteur | numéro lecteur. | Prénom | Nom | Lire () |
|---|---|---|---|---|
| numéro lecteur. | 1 | 0 | 0 | 0 |
| Prénom | 0 | 1 | 0 | 0 |
| Nom | 0 | 0 | 1 | 0 |
| Consulter () | 0 | 0 | 0 | 1 |

**Tableau 2: tableau de calcul de similarité entre les concepts**

On va s'appuyer sur l'ontologie de domaine pour constater que consulter() et lire() sont synonymes et par conséquent :σ (Consulter (), Lire()) = 1.
σ (OntoPersonne , OntoLecteur) = ¼ (1+1+1+1 + ∑0) =1

Les ontologies OntoPublication de Biblio1 et OntoPublication de Biblio2 ont la même apparence mais elles ont une mesure de similarité σ égale à 0, ce qui explique qu'il y a homonymie et risque de conflit de type nommage.

Conclusion : OntoPersonne et OntoLecteur sont synonymes ; OntoPublication de Biblio1 et OntoPublication de Biblio2 sont homonymes.

5. - Marquage des correspondances entre les concepts pour résoudre le problème des conflits sémantiques.
   - Création d'ontologies de représentation.
   - Détermination du CM résultat qui représente les connaissances des composants de départ.

A la fin de ce processus, nous avons obtenu un composant métier résultat qui encapsule les connaissances des autres composants. Ce nouveau composant métier peut alors être utilisé par les concepteurs pour réaliser le nouveau système d'information.

## 4. Conclusion et perspectives.

Notre proposition vise à permettre aux analystes et concepteurs de détecter les conflits sémantiques de type nommage entre composants métiers conceptuels, candidats à la réutilisation dans un projet système d'information. Elle consiste en une architecture pour l'intégration sémantique des composants métier, ainsi qu'une méthode de mesure de la similarité pour la détection et la résolution des conflits de type nommage. Notre solution est une application des ontologies au domaine de l'intégration des la conception des SI à base de composants. Les composants métiers ont représenté dans notre cas les ressources à décrire et à intégrer. Le champ d'application de notre solution concerne les composants métier conceptuels disponibles aussi bien dans un environnement centralisé que dans un environnement distribué.

Des exemples nous ont permis de vérifier la solution. Nous pensons poursuivre ce travail d'abord par une validation formelle de la solution, et ensuite par la recherche des possibilités de l'étendre pour résoudre les autres types de conflits sémantiques, notamment les conflits de mesure et de confusion.

## Références